\RequirePackage{fix-cm}
\RequirePackage{amsmath}
\documentclass{article}

\usepackage{graphicx}
\newtheorem{Theorem}{Theorem}[section]
\newtheorem{theorem}{Theorem}[section]
\newtheorem{definition}[Theorem]{Definition}
\newtheorem{lemma}[Theorem]{Lemma}

\newtheorem{proposition}[Theorem]{Proposition}

\begin{document}

\title{New Approaches to Principal Component Analysis for Trees
}


\author{Burcu Ayd{\i}n         \and
        G\'{a}bor Pataki \and
        Haonan Wang \and
        Alim Ladha \and
        Elizabeth Bullitt \and
        J.S. Marron
}



\date{}

\maketitle

\begin{abstract}
Object Oriented Data Analysis is a new area in statistics that studies populations of general data objects. In this article we consider populations of tree-structured objects as our focus of interest. We develop improved analysis tools for data lying in a binary tree space analogous to classical Principal Component Analysis methods in Euclidean space. Our extensions of PCA are analogs of one dimensional subspaces that best fit the data. Previous work was based on the notion of \emph{tree-lines}.

In this paper, a generalization of the previous tree-line notion is proposed: \emph{$k$-tree-lines}. Previously proposed tree-lines are $k$-tree-lines where $k=1$. New sub-cases of $k$-tree-lines studied in this work are the  \emph{$2$-tree-lines} and \emph{tree-curves}, which explain much more variation per principal component than tree-lines. The optimal principal component tree-lines were computable in linear time. Because $2$-tree-lines and tree-curves are more complex, they are computationally more expensive, but yield improved data analysis results.

We provide a comparative study of all these methods on a motivating data set consisting of brain vessel structures of $98$ subjects.
\end{abstract}

\section{Introduction}\label{Introduction}
The challenging problem of statistically analyzing samples drawn from populations of trees was first tackled by Wang and Marron (2007). Motivated by a data set of brain vessel structures, they developed an analog of the \emph{Principal Component Analysis} (PCA) technique in binary tree space. They replaced best fitting sub-spaces in PCA with best fitting tree-lines using appropriate definitions of distance, median, etc. in this new domain. They formulated a notion of principal components using these definitions.

Ayd{\i}n et al. (2009) gave linear time algorithms to calculate these principal components. Using these, they were able to conduct a numerical study on a motivating data set of brain artery structures of $73$ subjects from Aylward and Bullitt (2002).  This set was later further extended with more subjects and went through a data cleaning process as explained in Ayd{\i}n et al. (2011), resulting in an improved set which is used in the analyses conducted in this paper. 

The clinical findings of Ayd{\i}n et al. (2009), which resulted from the tree-line methodology, included a significant correlation between brain artery structure and age. They also were able to observe some symmetry properties across different regions of the brain.

While these results were promising, each tree-line principal component explained a quite small portion of the variation present in the set, due to the denseness of the data trees. In particular, no component gave much description of tree shape. This required the combination of many principal components to obtain a useful summary of the data. 

Our first contribution in this paper, the idea of \emph{$k$-tree-lines}, is a generalization which directly targets these drawbacks. In fact, the original tree-lines are the special case where $k=1$. The attractive aspect of $k$-tree-lines is that as $k$ increases, the possible shapes the components can take become more and more general. They allow more complex structures in principal components and promise richer results. However, this more complex structure also brings computational challenges. The linear time algorithm invented by Ayd{\i}n et al. (2009) for tree-lines motivated us to seek polynomial time algorithms for $k$-tree-lines.

In Section \ref{2treelines}, we show that a naive brute force calculation requires a high degree polynomial computational time using  a complexity argument, for $k=2$. We also develop a Branch and Bound ($B\&B$) algorithm to solve these problems, as well as numerical study results obtained using $2$-tree-lines.

Another special case we have examined is when $k=\infty$. The $\infty$-tree-lines consist of a sequence of trees in binary tree space where each tree is distance $1$ (in the sense of having one additional node) from the previous tree in the sequence. These sequences parallel curves in Euclidean space, and thus have been named \emph{tree-curves}. They provide the most general structure in the framework of $k$-tree-lines, and the richest numerical results. However, tree-curves are more challenging to compute. In fact no polynomial-time algorithm to compute the optimal tree-curves has been found by the authors. In Section \ref{Treecurves}, we introduce certain heuristics developed to find near-optimal results. Their results explain much more variation than was observed previously in the brain artery data by tree-lines. Moreover, they provide new insights about the underlying artery structure, such as structural differences between systems feeding different regions of the brain.

Other recent approaches to the statistical analysis of trees exist in the literature. See Banks and Constantine (1998) for a likelihood approach, Breiman et al. (1984) for classification and regression tree analysis, and Breiman (1996) and Everitt et al. (2001) for using trees in cluster analysis. As a more recent development, Nye (2011) provides a different approach to PCA in populations of trees within the phylogenetic trees context.

There are also other studies that specifically focus on analysis of binary trees, and apply findings to brain artery data. For example, Bullitt et al. (2010) uses average node number of each tree as a summary statistic. This method does not capture any shape-related aspect, but can relate the overall size of the trees to an external parameter. Wang et al. (2011) gives a nonparametric regression model for tree shaped data, and Alfaro et al. (2011) develops a dimension reduction technique for PCA in trees. Shen et al. (2011) takes the Dyck path formulation approach to this problem and employ functional data analysis methods.

\subsection{Data Description and Tree Representation}

The properties of the motivating data set and the extraction of binary trees from the $3D$ brain vessel images are explained in Ayd{\i}n et al. (2009) in detail. Here we will give a brief summary for the sake of completeness.

\begin{figure}
[ptb]
\begin{center}
\includegraphics[natheight=1.4in,natwidth=2.1in,height=1.4in,width=2.1in]{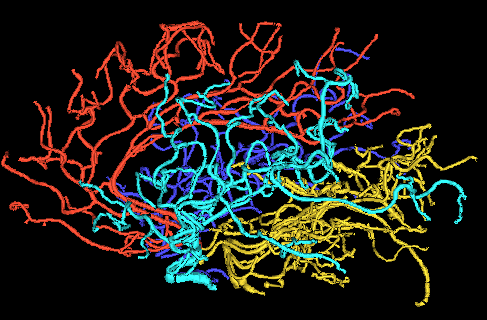}%
\includegraphics[natheight=1.4in,natwidth=2.1in,height=1.4in,width=2.1in]{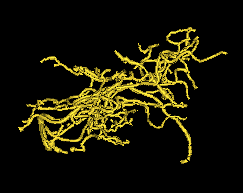}%
\includegraphics[natheight=1.4in,natwidth=2.1in,height=1.4in,width=2.1in]{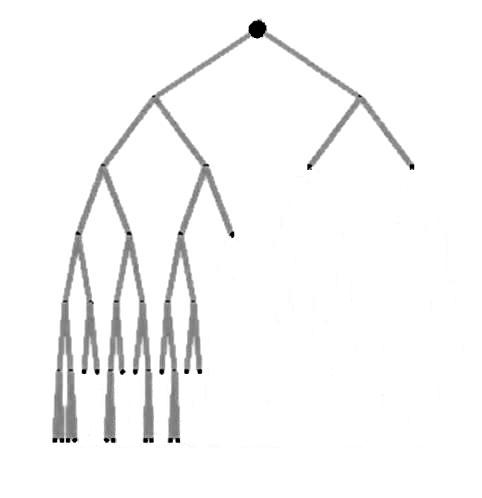}%
\caption{Left panel: Reconstructed set of trees of brain arteries. The colors indicate regions of the brain: Back (gold),
Right (blue), Front (red), Left (cyan). Middle Panel: The back sub-system is shown only. Right panel: Binary tree obtained from the Back tree (gold) of the same subject. Only branching information is retained.}
\label{Fig1}%
\end{center}
\end{figure}

The data are from a Magnetic Resonance Angiography (MRA) study of brain images of a set of $98$ human subjects of both sexes, ranging in age from $18$ to $72$, which can be found at Handle (2008). A tube tracking algorithm was applied to the MRA images resulting in a segmentation of arteries as shown in the $3D$ images in Figure \ref{Fig1}. See Aylward and Bullitt (2002) and Bullitt et al. (2010) for details of this study.

The artery system feeding the brain can be divided into $4$ component systems according to the areas they feed in the brain. In the figure, these systems are colored in gold for the back, cyan for the left, blue for the right and red for the front regions. Each of these regions are studied separately, giving rise to $4$ data sets. For each of these regions, the $3D$ vessel structure is reduced to only its topological (connectivity) aspects by representing it as a simple binary tree. Each vessel tube between two split points is converted into a node in the binary tree, and the two tubes after the split are the children nodes of the first node. Figure \ref{Fig1} gives an example of this conversion. The root node at the top represents the initial fat gold tree trunk shown near the bottom of the figure.

There is one ambiguity in the construction of the representation shown in the right panel in Figure \ref{Fig1}. That is the choice, made for each split, of which
child branch is put on the left, and which is put on the right. The word \emph{correspondence} is used to refer to this choice. Throughout this paper we will use the \emph{descendant correspondence}, where the child with the most number of descendants is assigned to be the left child.

Statistical analysis of the brain artery data is important in understanding how various factors affect this structure, and how they are related to certain diseases (as noted below). In this paper, the connection between aging and branching structure is the main focus. This connection was previously explored in studies such as Ayd{\i}n et al. (2009) and Bullitt et al. (2010). Bullitt et al. (2010) identified that the number of brain vessels observable by MRA decreases with age in healthy subjects. Ayd{\i}n et al. (2009) tied these effects to structural properties. For a detailed account of vascular changes observed in the brain and its ties to aging, the reader is referred to Bullitt et al. (2010).

Apart from the discussion of aging effects, the study of brain vessel structure is important as it is thought to be related to hypertension, atherosclerosis, retinal disease of prematurity, and with a variety of hereditary diseases. Furthermore, there is thought to be a causal relationship between vessel structure and thrombosis or stroke. Therefore results of studying this structure may lead to establishing ways to help predict risk of these diseases. Another very important implication regards malignant brain tumors. These tumors are known to change and distort the artery structure around them, even at stages where they are too small to be detected by conventional imaging techniques. Statistical methods that might differentiate these changes from normal structure may help earlier diagnoses. See Bullitt et al. (2003) and the references therein for detailed medical studies focusing on these subjects.

Our numerical analysis in this paper solely focuses on the brain vessel analysis. However, the statistical tools proposed in this paper are applicable to any binary tree data set where the statistical trends are of interest. Some examples include other vessel structures in the body, lung airway systems, plant root development systems, and organization structures. In fact, Alfaro et al. (2011) apply their backward PCA for trees to investigate the properties of the organization structure of a large company.

Our models in this paper are based on the definitions of the binary tree and a distance metric given in Wang and Marron (2007). A \textbf{binary tree} is a set of nodes that are connected by edges in a directed and acyclic fashion, which starts with one node designated as \textbf{root}, where each node has at most two children. Using the notation $t_{i}$ for a single tree, let:
\begin{equation*} \label{Tti}
T=\left\{  t_{1}%
,...,t_{n}\right\}
\end{equation*}
denote a data set of $n$ such trees. Given two trees $t_{1}$ and $t_{2}$, their  (Hamming) \textbf{distance} is
\[
d\left(  t_{1},t_{2}\right)  = | t_{1}\backslash t_{2} | + |  t_{2}\backslash t_{1} |,
\]
where $\backslash$ denotes set difference and $|.|$ denotes the cardinality of the set. The union of all data trees in a given set is defined to be the \emph{support tree} ($Sup(T)=\cup_{i=1}^n t_i$).

\section{Formulation of $k$-Tree-Lines}\label{Ktreelines}

The idea of \emph{$k$-tree-lines} is developed as a generalization of the tree-line concept, in an attempt to overcome the limitations of tree-lines and provide the analyst with a set of tools capable of examining data from various angles. When constructing a tree-line, each tree is obtained by adding a child to the last node of the previous tree. This last node, whose children are candidates for addition, is called an \emph{active} node. In a $k$-tree-line, at each step, the $k$ nodes that were added last are active.
The formal definition of a $k$-tree-line is:

\begin{definition}
A\textbf{\ $k$-tree-line}, $K=\left\{  \ell_{0},\cdots,\ell_{m}\right\}  $, is a sequence of trees where $\ell_0$ is called the starting tree, and $\ell_{i}$ comes from $\ell_{i-1}$ by the addition of a single node, labeled $v_{i}$. \ In addition, each $v_{i+1}$ is a child of one of the nodes in $\{v_{i-k+1},\cdots,v_i\}$, or in the case where $k>i$, it is a child of one of the members of $\{\ell_0,v_1,\cdots,v_i\}$. A $k$-tree-line of which the last $k$ nodes are leaves of the support tree, that is, a $k$-tree-line that cannot be further extended is called a \textbf{maximal} $k$-tree-line. All other lines are called \textbf{partial} $k$-tree-lines.
\end{definition}

It can be seen that the $k$-tree-line is a generalization of the previously proposed tree-line structure, which is now $k=1$. Higher order $k$'s are useful because for lower orders, such as $k=1$, each individual covers only a small region of the tree space. In the limit as $k\rightarrow\infty$, this structure becomes a tree-curve, as detailed in Section \ref{Treecurves}.

A key concept to develop a principal component analysis framework is the idea of projection. In the most general sense, the projection of a data point $t$ onto an object or subspace, essentially a set of points living in the same space as $t$, is the point(s) in that set that have the smallest distance to $t$. Extending this general concept to our case, the projection of a data tree onto a $k$-tree-line is a point on the $k$-tree-line with smallest distance to the data tree:

\begin{definition}
Given a data tree $t$, its \textbf{projection} onto the $k$-tree-line $K$ is%
\[
P_{K}\left(  t\right)  =\underset{\ell \in K}{\arg\min} \{ d\left(t,\ell \right)\}.
\]
\end{definition}

Unlike the tree-line case, the projection of a data point does not have to be unique.

Similarly, one can extend the general idea of principal components into the $k$-tree-line structure. In Euclidean space, a principal component of a given data set, is a one dimensional sub-space (line) that minimizes the sum of squared distances between data points and their projections. This is extended to $k$-tree-lines as:

\begin{definition} \label{def-kl1star}
For a data set $T$, the \textbf{first principal component $k$-tree-line} is
\[
K_1^* \, = \, \underset{K}{\arg\min} \sum_{t_i \in T} d(t_i,P_{K}(t_i))
\]
\end{definition}


This definition is extended to additional principal components as:

\begin{definition}\label{something}
For $j > 1$ the \textbf{$j$th principal component $k$-tree-line} is defined recursively as:
\[
K_j^*= \underset{K}{\arg\min} \sum_{t_i \in T} d(t_i,P_{K_1^*\cup \cdots\cup K_{j-1}^*\cup K}(t_i))
\]
\end{definition}

It was shown in Claim $3.1$ of Ayd{\i}n et al. (2009) that the optimal principal components for $1$-tree-lines are maximal, and the projection of a data point onto them is unique. For cases $k>1$, the uniqueness of projection is not guaranteed.\footnote{For $k=2$, a simple counter-example where projection is not unique can be constructed. By definition, the set of $k_1$-tree-lines include the set of $k_2$-tree-lines if $k_1 \geq k_2$. Therefore the non-uniqueness trivially extends to all $k>1$.} However, the set of optimal solutions to the best principal components problem for $k>1$ contains at least one maximal $k$-tree-line, therefore maximality can be maintained.

What is provided so far is the adaptation of classical PCA ideas to the $k$-tree-line structure. Although there is one single generic formulation of principal components for all $k$-tree-lines, each $k$ yields a very different optimization problem. The next two sections will focus on the cases where $k=2$ and $k=\infty$.

\section{Study of $2$-Tree-Lines}\label{2treelines}

\subsection{A Complexity Argument}\label{compl}

The first step in solving the $2$-tree-line problem is to determine if a polynomial-time solution exists.

\begin{lemma} \label{cl:2lineorder}
For a data set with a full support tree of $m$ nodes, the number of all $2$-tree-lines within its support tree has an order of $O(m^{2.9})$.
\end{lemma}

The proof of Lemma \ref{cl:2lineorder} is in the Appendix. This result is used to obtain the following theorem:

\begin{theorem}\label{cl:2linethm}
For a data set with a full support tree of $m$ nodes, the run time of the brute force method of checking all possible $2$-tree-lines has an order of $O(m^{2.9}\log m)$.
\end{theorem}

The proof of Theorem \ref{cl:2linethm} can also be found in the Appendix.

Theorem \ref{cl:2linethm} establishes that we have a polynomial time problem. While the polynomial bound is promising, we can get faster convergence using a B\&B based algorithm.



\subsection{Solution Methods}\label{2methods}

The method we propose in this paper to quickly solve $2$-tree-line problems is based on a partition based strategy called Branch and Bound. B\&B refers to a wide range of algorithms used to solve global optimization problems. The method was first proposed in Land and Doig (1960). The approach is especially useful when a convex feasible region structure is not available: Such as in integer programming (Schrijver (1998)), various combinatorial optimization problems (Cook et al. (1997)), and nonlinear programming (Bazaraa et al. (1979)). For a general introduction, Lawler and Wood (1966) and Lawler and Bell (1966) provide a good starting point for the interested reader.

\subsubsection{The Generic Branch \& Bound Method}\label{genericBB}

Consider the general optimization problem $\mathcal{OP}$:
\[
\mbox{Minimize} \ \ G=g(x) \ \ \mbox{Subject to:} \ \ x \in \mathcal{F}\\
\]
where $\mathcal{F}$ represents the set of all feasible solutions to the problem $\mathcal{OP}$, and $G^{*}$ is the optimal solution value being sought. Notice that the definition of $\mathcal{OP}$ is generic enough so that almost any optimization problem can be written in this form.

Let the set $\mathcal{F}$ be partitioned as follows: $\mathcal{F}=\mathcal{F}_1 \cup \ldots \cup \mathcal{F}_n$, where the subsets $\mathcal{F}_i$ are disjoint. Then for each $i$ define the sub-problem $\mathcal{OP}_i$ as:

\[
\mbox{Minimize} \ \ G_i=g(x) \ \ \mbox{Subject to:} \ \ x \in \mathcal{F}_i
\]

Note that:

\[
G^* = \min_{i = 1 \ldots n}\{G_i^*\}
\]

This process of partitioning a bigger problem into smaller portions is called \emph{branching}. The B\&B algorithm is an iterative process that partitions existing subproblems into smaller subproblems at each step. It follows these steps at each iteration:

\begin{enumerate}
  \item Determine the current partition.
  \item Recognize if the problem $\mathcal{OP}_i$ has no feasible solution and thus $\mathcal{F}_i = \emptyset$, if not, find a feasible point $x^{feas}_i \in \mathcal{F}_i$.
  \item Solve a \emph{relaxed} version of the problem $\mathcal{OP}_i$, and obtain a point $x^{rel}_i$. This point may or may not be in $\mathcal{F}_i$.
  \item For all non-empty partition pairs $i$ and $j$, check whether $g(x^{feas}_i) < g(x^{rel}_j)$. If this holds, remove $\mathcal{F}_j$ from consideration.
\end{enumerate}

The last step of finding bounds for each subproblem is called \emph{bounding}, and the removed infeasible or dominated subproblems (or branches) are called \emph{cut} or \emph{pruned}.

For each of the subproblems with a nonempty feasible region, the following holds:

\[
G_i^* \in [g(x^{rel}_i),g(x^{feas}_i)]
\]
At any point during the progress of the algorithm, it is known that:
\[
G^*  \in [\underset{i}{\min}\{g(x^{rel}_i)\},\underset{i}{\min}\{g(x^{feas}_i)\}]
\]
This bracket is called the \emph{optimality gap}. The B\&B algorithm terminates when a point in $\mathcal{F}$ gets singled out as the optimal solution, or when a sufficiently small optimality gap is reached.

\subsubsection{Adaptation to $2$-Tree-Lines}\label{adopt}


The $2$-tree-line adaptation involves defining appropriate partitions of the feasible region of all possible $2$-tree-lines in $Sup(T)$ given a data set $T$ and a starting point $\ell_0$. Due to the nature of the algorithm, at each iteration, there are two sets generated: One containing the new candidate set ($\mathcal{K}$) of the step, and another set ($\mathcal{C}$) generated by applying the pruning action to $\mathcal{K}$. The active feasible region at each iteration consisting of all the still possible maximal $2$-tree-lines also needs to be explicitly defined ($\mathcal{F}$). Therefore, to develop the B\&B algorithm for $2$-tree-lines, it is useful to define the following three sequences of sets:

\begin{definition}\label{partials}
Using $\mathcal{K}$ to denote the current set of active partials,  let $\mathcal{K}^0 = \{\ell_0\}$. $\mathcal{K}^n$ is the set of all partial $2$-tree-lines that can be obtained by adding one node to the partial $2$-tree-lines contained in $\mathcal{C}^{n-1}$. $\mathcal{C}^n$ is the set of partial $2$-tree-lines remaining after the pruning step of the B\&B algorithm is performed on $\mathcal{K}^n$. The set of all maximal $2$-tree-lines that can be obtained by extending the $j^{th}$ member of $\mathcal{K}^n$ is $\mathcal{F}^n_j$, and $\mathcal{F}^n=\underset{j}{\bigcup}\mathcal{F}^n_j$.
\end{definition}

Clearly, $\mathcal{F}^0$ corresponds to the feasible region of our initial problem. At each step $n$, $\mathcal{F}^n$ is the union of the active partitions at that step.

Determining whether a set $\mathcal{F}^n_j$ is empty, and finding a maximal $2$-tree-line that includes a member of $\mathcal{F}^n_j$ (called $x^{feas}_j$) are rather trivial. However, choosing an $x^{feas}_j$ that will provide a tighter upper bound will improve the convergence of the algorithm.

The task of defining and solving a relaxation of the problem requires more attention. For this, we will first introduce the definition of \emph{weight}, and \emph{$2$-path}:

\begin{definition}\label{weight}
Given a data set $T$, the \textbf{weight} of a node $v$ is the number of times it occurs in the set $T$:
\[ w(v)= \sum_{t_i\in T}\delta(v,t_i)\]
\end{definition}

A useful lower bound in the B\&B problem can be provided by the following:

\begin{definition}\label{2path}
A \textbf{$2$-path} is a rooted tree which includes at most $2$ nodes at each level. A $2$-path of a $2$-tree-line $K$ is the smallest $2$-path that contains all the members of $K$ and is denoted as Pa($K$). The \textbf{maximum $2$-path} of a partial $2$-tree-line $K$ in a support tree $Sup(T)$ is the $2$-path with maximum sum of weights that is contained in $Sup(T)$, and contains all members of $K$. It is denoted as MP($K$).
\end{definition}

The solution of the \emph{maximum 2-path problem} for any partial $2$-tree-line can be used as a lower bound in the B\&B algorithm:

\begin{proposition}\label{lowerbound}
For a given partial $2$-tree-line $K$, $\sum_{v \in MP(K)}w(v)$ provides a lower bound on the best maximal $2$-tree-line that can be extended from $K$.
\end{proposition}

The proof of Proposition \ref{lowerbound} can be found in the Appendix.

A dynamic programming approach is used to find the maximum $2$-path of the active partial $2$-tree-lines at each step. 

For any region $\mathcal{F}^n_j$, any feasible point (any maximal $2$-tree-line) within the region can be used to obtain an upper bound. However, a tight upper bound can be reached if a $2$-tree-line that contains the maximum $2$-path of the region is used. The numerical results at the end of the section verify that these lines indeed provide very close, if not exact, approximations of the objective function value, increasing the convergence of the algorithm dramatically.





Under the light of these, a step by step description of the $2$-tree-line B\&B algorithm can be given as follows:

\emph{Inputs:} $T=\{t_1,t_2,...,t_n\}$ is the binary tree data set, and $\ell_0$ is the starting tree.


For each $i$:
\begin{itemize}
  \item Form the set $\mathcal{K}^i$ by extending each of the partial $2$-tree-lines in $\mathcal{C}^{i-1}$ with all possible next nodes.
  \item For each $K \in \mathcal{K}^i$:
       \begin{itemize}
       \item Determine $MP(K)$, and a maximal $2$-tree-line $K^{max}$ that passes through it.
       \item Calculate the lower bound $LB^K=\sum_{t_i \in T}{|t_i|} - \sum_{v \in MP(K)}{w(v)}$ and the upper bound $UB^K=\sum_{t_i \in T}d(t_i,P_{K^{max}}(t_i))$ for this partition.
       \end{itemize}
  \item For any partial $2$-tree-line pair $\{K,J\}$, if $UB^J < LB^K$, then partial $K$ is dominated by partial $J$, so delete $K$ from the list. Obtain $\mathcal{C}^i$ by deleting all dominated partial $2$-tree-lines.
\end{itemize}
Stop when a set of optimal maximal $2$-tree-lines are reached. The output list of $2$-tree-lines obtained have the same upper and lower bounds since they are maximal lines, and thus they have the same objective function values.



\subsection{Performance Analysis of the $2$-Tree-Line B\&B} \label{sec:kperf}

In this section, first, we will introduce a simulation study that compares the performance of the $B\&B$ algorithm to that of the naive brute-force method. Second, we will show the performance of the $B\&B$ on the real data set.

In terms of performance analysis, there are several measures that can be used to determine the contribution of an algorithm to computational power. One can investigate the size of the largest problem instance that can be solved with previous methods and compare it with the possible size that the new method can deal with. Another possibility is to compare computation times of previous and new methods for the same instances.

To illustrate the performance differences, we created $100$ data sets, each data set consisting of $10$ random data trees. To create each of the binary data trees, we assume that each of the nodes either branch into $2$ children with probability $p$, or do not branch and therefore remain a leaf node with probability $1-p$. Each data tree contains at least the root node.

The system we use to denote the nodes comes from Wang and Marron (2007), where a unique integer is used to denote each possible location for a node. These integers have a potential to get very large in deeper levels of a tree. The mathematical program we employ for this work, MATLAB $R2011b$, only stores values up to $2^{53}-1$ for double variables. This allows for trees at most $53$ levels deep. To avoid numerical issues, we limit the size of our simulated data trees to at most $53$ levels.

For a given binary tree where nodes either branch into $2$ children or do not branch, if it is assumed that every node has the same branching probability $p$, then this $p$ can be estimated from the formula $\hat{p}=\frac{1}{2}(1-\frac{1}{n})$, where $n$ is the size of the tree. We have calculated the estimated branching probability $\hat{p}$ for all our data trees in the brain artery set, and used the average of it ($0.4953$) to create the simulated data trees. This group of $100$ random data sets will be called $SET1$.

The trees in $SET1$ branch completely randomly, and there is no underlying trend in these sets. For real-life data sets, this is usually not the case. For example, the brain artery data set consists of trees that carry a lot of structural similarities with each other. Some of the similarity is coming from the descendant correspondence. This allowed for making sure the nodes representing the larger arteries align across data trees. A consequence of this correspondence is left-heavy data, which naturally carries a high level of common structure within itself. To mimic this common structural trend, the trees in $SET1$ are re-arranged according to the descendant correspondence to form $SET2$. The size of each data tree does not change after this procedure, but the common structure introduced reduces the size of the support trees.

We ran the $B\&B$ algorithm and the brute force method on all of the data sets in $SET1$ and $SET2$. The implementations were done in MATLAB $R2011b$. A personal computer with $2.53$ GHz Intel processor and $4$ GB RAM running $64$-bit Windows $7$ operating system is used for all the runs.

Some data sets happen to contain very large trees that may cause very large run times or may lead to memory problems. To manage the run-times, we set an upper limit of $500$ seconds for each of the data sets and methods. That is, both of the methods are set to terminate when an upper limit of $500$ seconds is reached. This time limit is long enough to compute the $2$-tree-line PC's for reasonably sized data sets. However, it is not long enough to allow the algorithms to reach memory limitations, therefore memory limits are not studied in this simulation.

\begin{figure}
[ptb]
\begin{center}
\includegraphics[scale=0.45]{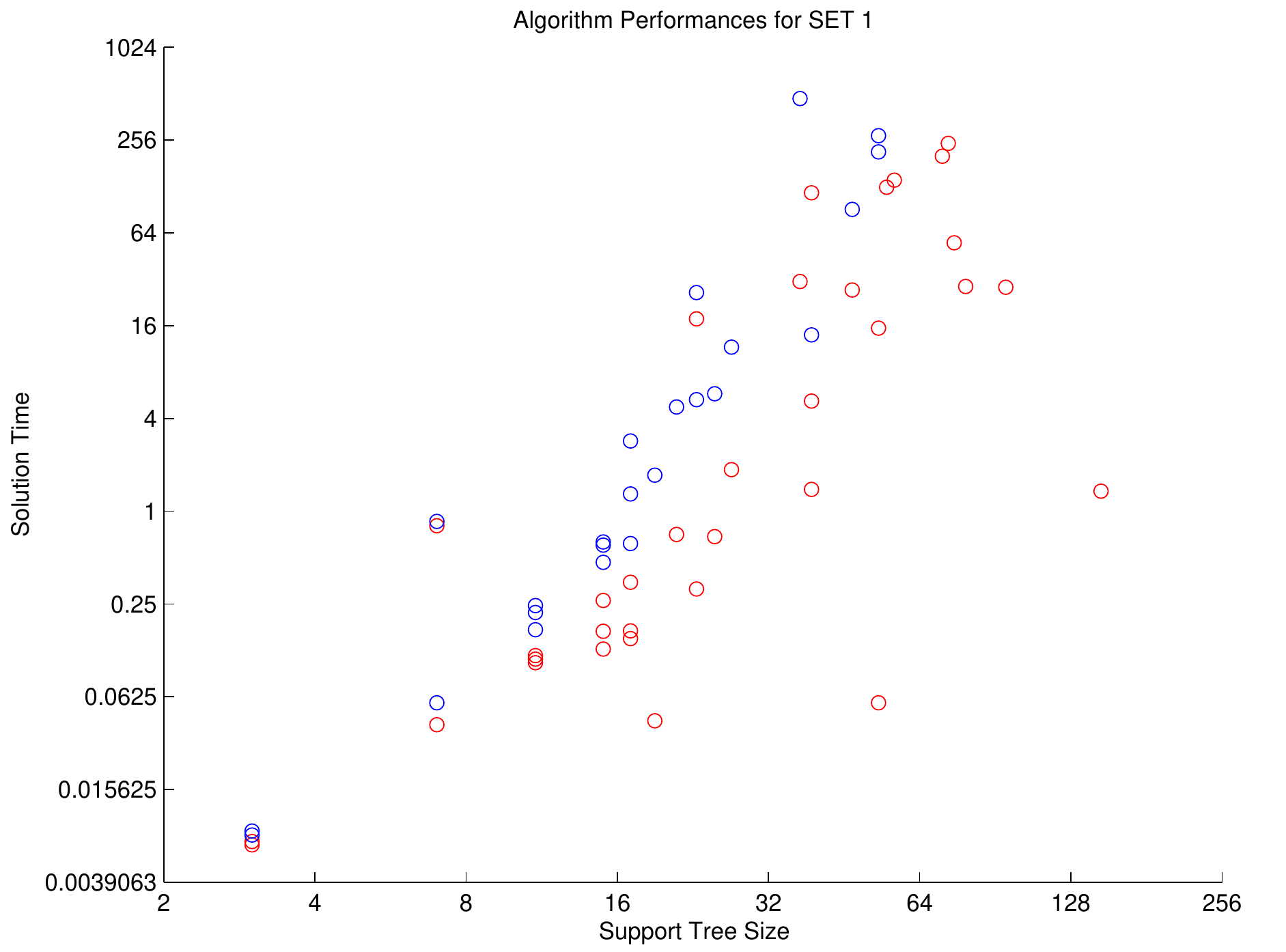}
\includegraphics[scale=0.45]{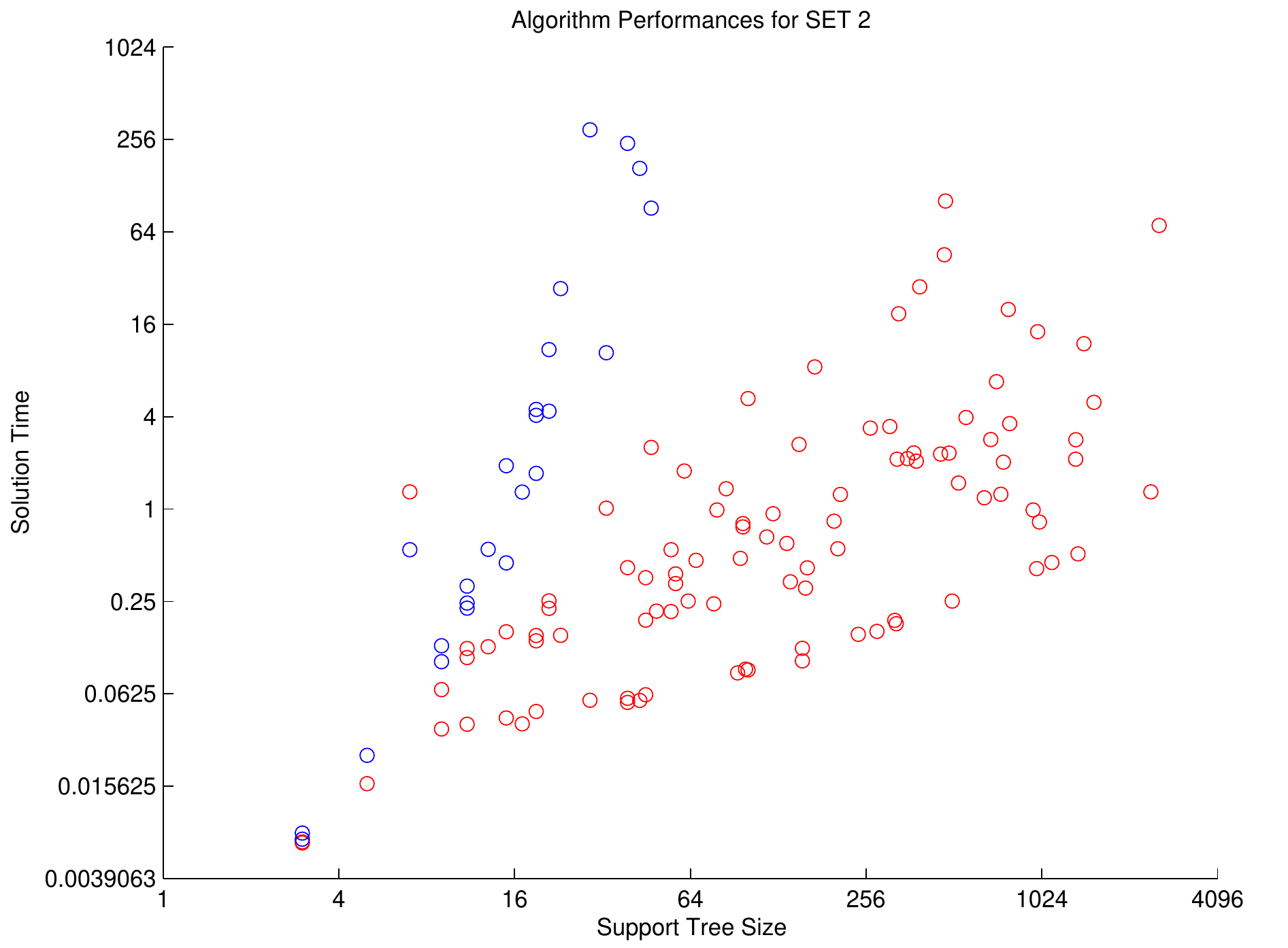}
\caption{Graphs showing the run times of $B\&B$ algorithm (red) and brute force algorithm (blue) for the data set instances for which they were able to reach the optimal solution within $500$ seconds. Upper panel is results of $SET1$, lower panel is results of $SET2$. The $X$ coordinates are the support tree sizes for the data sets, shown on a log scale. $Y$ coordinates are the run times, also on a log scale. The axis labels are given in actual seconds and sizes. For $SET2$, $B\&B$ is dramatically better.}
\label{sim_run_time}
\end{center}
\end{figure}

The run time of both of the algorithms depends on various aspects of the data set. These include the shape of the data trees and the sizes of them. The size of the support tree of a data set can be a good indicator of problem difficulty, although it is not the sole indicator. Figure \ref{sim_run_time} shows the solution times obtained by both of the methods for each data set versus the size of the support trees of these data sets.

We focus our study on the data sets for which either of the methods can find an optimal solution within the allotted time of $500$ seconds. For $SET1$, out of the $100$ data sets, the $B\&B$ found the optimal solution for $34$ of them, and brute force method found the optimal solution for $24$ of them. None of the data sets in this trial had the optimal solution found by brute force method but not by $B\&B$.

For the $24$ data sets that both of the algorithms reached the optimal solution, the average solution time for the brute force method was $47.2$ seconds, whereas it was $4.1$ seconds for the $B\&B$ algorithm. The largest data set, for which the brute force method could find the optimal solution, has support tree size of $53$ nodes. The $B\&B$ algorithm could find the optimal solutions for up to $147$-node support trees.

For $SET2$, the brute force method reached the optimal for $24$ instances, and $B\&B$ found the solutions for $98$ of the data sets within the $500$ seconds. Out of the $24$ instances solved by both methods, the average solution time was $35.97$ seconds for the brute force method, and $0.29$ seconds for the $B\&B$. The largest set for which the $B\&B$ successfully found a solution had $2585$ nodes in its support tree, while the largest set to be solved by brute force had $47$ nodes.

Overall, the simulation results show that the $B\&B$ algorithm greatly improves the run times needed to find the optimal solution, and it enables the analysis of larger data sets. The comparison of $SET1$ and $SET2$ shows that, $B\&B$ provides significant improvements over the naive method even when the branching structure is completely random ($SET1$). However, the real difference is observed when a common structure is introduced to the data sets ($SET2$). $B\&B$ takes advantage of this by quickly eliminating the more unlikely solutions early on, while the brute force method does not differentiate between these. This capability allows $B\&B$ to solve very large instances in within small amounts of time.

\begin{figure}
[ptb]
\begin{center}
\includegraphics[natheight=2.5in,natwidth=4in,height=2.5in,width=4in]{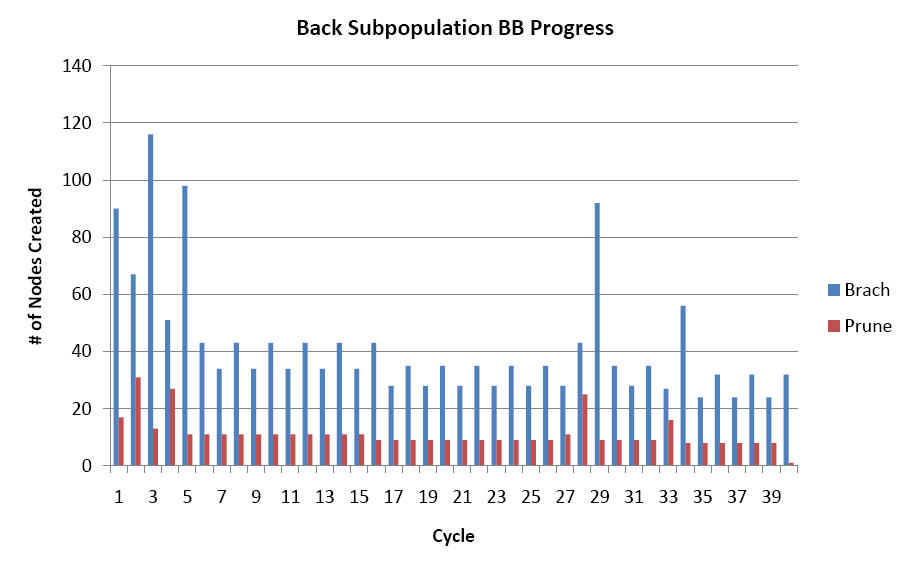}
\caption{Graph showing number of partial lines considered by the B\&B algorithm for the Back sub-population. Blue bars indicate the number of partial lines created at the beginning of each step. Red bars give the number of partial lines remaining after the pruning step for each iteration. Note that this number remains small throughout the algorithm progress.}
\label{BBprogress}
\end{center}
\end{figure}

Figure \ref{BBprogress} summarizes the progress of B\&B for each of the Back sub-population. The $x$ axis indicates each iteration and the length of the $x$ axis shows the number of iterations run before the optimal value is reached. The $y$ axis is on the scale of number of partial lines. The blue bars indicate the number of partial lines created at that iteration ($|\mathcal{K}_i|$), while the red bars give the number of remaining partial lines at that iteration after the pruning step is executed ($|\mathcal{C}_i|$). The graphs for the remaining sub-populations are very similar to this one, and therefore omitted from the text.

The size of the largest problem that can be handled by the brute force method has not been measured, but experience revealed that the optimal $2$-tree-lines for current data sets could not be found using the previously mentioned personal computer. The memory requirement for the number of $2$-tree-lines that need to be stored for these data sets seems to exceed the current capacity. The $B\&B$ algorithm terminates in $O(\log n)$ steps for a data set with a full support tree of size $n$. As seen in Figure \ref{BBprogress}, the largest number of partial lines that needs to be stored by the $B\&B$ algorithm at once is $131$, therefore the memory limitation has been overcome.



\subsection{Analysis of the Brain Artery Data} \label{sec:kanal}

One interesting question regarding the $2$-tree-lines is, how much of the existing variation in the data sets can they explain compared to the $PC1$ and $PC1\cup2$ calculated from the earlier $1$-tree-lines? It is reasonable to expect the coverage of $PC1\cup2$ of $1$-tree-lines to be close to the coverage of the first $2$-tree-line. 

\begin{table}[ptb]
\begin{center}
\begin{tabular}{| c | c | c | c | c |}
\hline
             & Back    & Left     & Right   & Front \\ \hline
$PC_11$      &2501 (18\%)& 2449 (22\%)& 2633 (22\%)& 2336 (25\%)\\
$PC_11\cup2$ &3039 (22\%)& 2817 (25\%)& 3008 (26\%)& 2749 (29\%)\\
$PC_21$      &3336 (24\%)& 3232 (28\%)& 3404 (29\%)& 3006 (32\%)\\
$PC_21\cup2$      &4412 (32\%) &3968 (35\%)& 4154 (35\%)& 3832 (40\%)\\
\hline
\end{tabular}
\caption{The number of nodes explained by $PC_11$, $PC_11\cup2$, $PC_21$ and $PC_21\cup2$. The percentages of these relative to the total number of nodes are given in parenthesis. Note that $PC_21$ always explains more than $PC_11\cup2$.}
\label{scoretab}
\end{center}
\end{table}

Table \ref{scoretab} shows the number of nodes explained by the first $1$-tree-line PC ($PC_11$), the combination of the first and second $PC$'s of the $1$-tree-lines ($PC_11\cup2$), the first $2$-tree-line PC ($PC_21$), and the combinations of the first and second PC's of $2$-tree-lines ($PC_21\cup2$) for all four sub-populations. The percentages of these to the total number of nodes are given in parenthesis. The score of $PC_21$ is consistently higher than that obtained by $PC_11\cup2$. This tells us that the first $2$-tree-line explains more than the first two $1$-tree-lines combined.

The second question is: What information can we infer about the underlying structure of our data sets using $2$-tree-lines? The first principal components of $1$-tree-lines provided valuable insight on symmetry issues. Now we will investigate if the same observations are available using the $2$-tree-line analysis and if any more insights can be obtained.

\begin{figure}
[ptb]
\begin{center}
\includegraphics[scale=0.5]{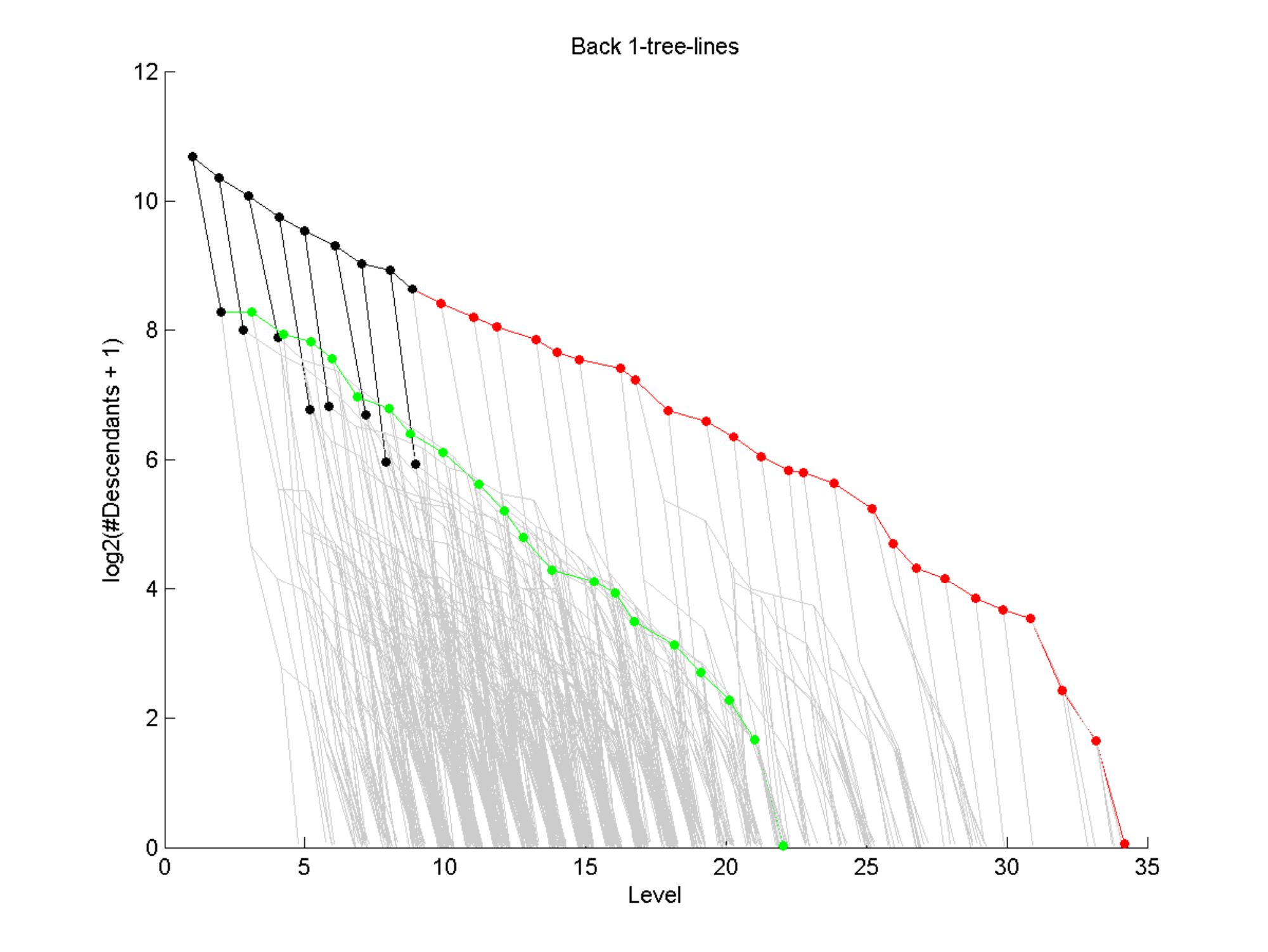}
\includegraphics[scale=0.5]{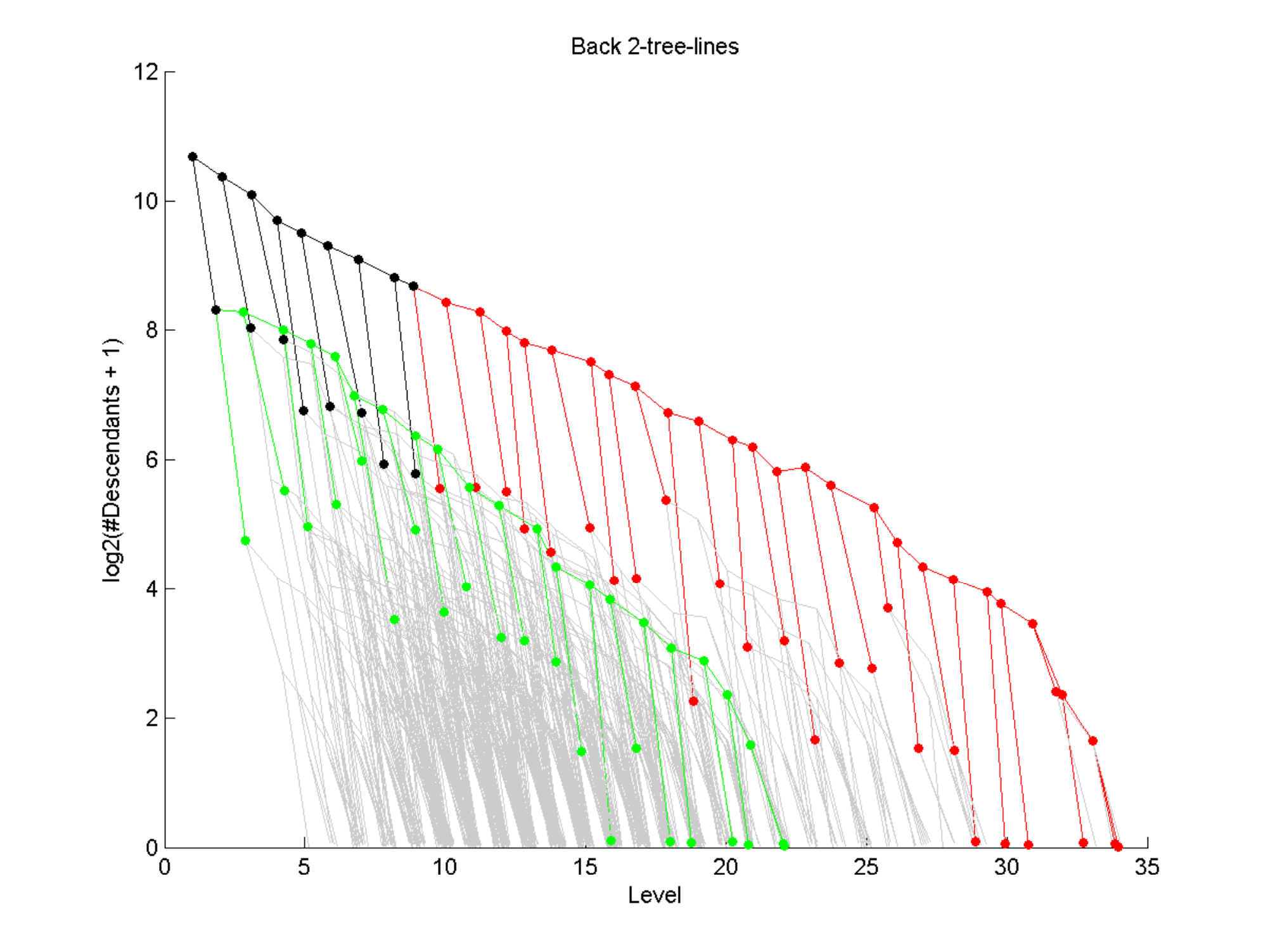}
\caption{Comparison of $1$-tree-lines and $2$-tree-lines. On the top, $PC_11$ and $PC_12$, on the bottom, $PC_21$ and $PC_22$ for the Back sub-population. The black nodes represent the starting point data tree, red nodes indicate the first principal component, and green nodes indicate the second principal component.}
\label{backlines}
\end{center}
\end{figure}

Figure \ref{backlines} depicts the first two $2$-tree-lines and the first two $1$-tree-lines drawn on the Back sub-population's support trees. The visualization technique used to produce these images is explained in detail in Ayd{\i}n et al. (2011). The \emph{D-L view} is developed to display large trees in limited space. Each node is located such that its $X$-coordinate is the level of the node in the binary tree ($1$ corresponding to the root level) and its $Y$-coordinate is the base-$2$ logarithm of that node's number of descendants. The nodes are connected according to their parent-child relationships.

In Figure \ref{backlines}, the black nodes indicate the starting trees in all plots, while red nodes constitute the first principal components ($PC_11$ on the top side and $PC_21$ on the bottom side plots) and green nodes are the second principal components ($PC_12$ on the top side and $PC_22$ on the bottom side). The right, left and front sub-populations present very similar pictures and are omitted here.

The principal components of the $2$-tree-lines follow the path of principal components of the $1$-tree-lines, with the exception that siblings of the same nodes now appear on the line. This is a consequence of the construction scheme of the binary trees from the original $3D$ images. In the original images, whenever a vessel split into two smaller vessels, two nodes are added to the corresponding binary tree, and thus two sibling nodes on the binary tree represent the trunks of two vessels that split from one parent vessel trunk. Therefore the binary trees in the data sets have nodes with either zero or two children. In other words, if a node exists in one of the binary trees, then so does its sibling.

The $1$-tree-lines can only follow a $1$-path in the support tree, therefore the double-node nature of the data sets is lost. $PC_11$ follows the path determined by the sibling which is the parent of the rest of the nodes on the line. Although each of the nodes on the line has a sibling with the exact same weight, they cannot appear on $PC_11$ due to the structural limitation, and $PC_12$ simply follows another path instead of covering these sibling nodes since its nodes have to be connected. The $2$-tree-lines seem to remedy this shortcoming. The same pattern is observed between $PC_12$ and $PC_22$.

This reasoning also explains why the $PC_21$ explains more nodes than $PC_11\cup2$. The $PC_11$ goes through the path with maximum sum of weights, and $PC_12$ through a path that has a slightly smaller sum. The siblings of the nodes on the $PC_11$ path also have the same exact weight count, so being able to include them into the $PC_21$ results in a better coverage than $PC_11\cup2$. Note that the score of $PC_21$ is not the double of $PC_11$ in Table \ref{scoretab} since the starting tree also contributes to the scores.

Finally, the age effect on the $2$-tree-line scores is investigated. It was previously shown that, there is a negative correlation between the ages of healthy subjects and the total number of vessels in their brains observable by MRA (Bullitt et al. (2010)). The first PCA analysis of trees enabled the researchers to summarize the structural trends in vessel systems, and observe the effect of aging on these summary trends rather than the whole data set. Ayd{\i}n et al. (2009) showed this effect using $1$-tree-lines. In this section, we will investigate the same effect using the $2$-tree-line tool, which has richer representation capabilities.

To do this, a simple linear regression is run for each case, where the predictor is the size of projections of each data point onto the principal components (scores), and the response is age. In other words, we investigate how size of the $2$-tree-line projections of data points are related to age.
The fitted regression lines have a negative slope, indicating lower scores may be associated with older ages. We test this observation against the null hypothesis of zero slope (no relationship). The slope $p$-values for all sub-populations are listed in Table \ref{pvaltab}, along with the slope $p$-values obtained from the $1$-tree-line principal components.

\begin{table}[ptb]
\begin{center}
\begin{tabular}{| c | c | c | c | c |}
\hline
             & Back & Left & Right & Front \\ \hline
$PC_11$      &0.0156& *    & 0.0186& * \\
$PC_11\cup2$ & *    & *    & 0.0002& * \\
$PC_21$      &0.0159& *    & *     & * \\
$PC_21\cup2$ & *    & *    & 0.0113& *  \\
\hline
\end{tabular}
\caption{The slope $p$-values obtained by $PC_11$, $PC_11\cup2$, $PC_21$ and $PC_21\cup2$ for all sub-populations. The slope $p$-values above the $0.05$ significance limit are marked with (*).}
\label{pvaltab}
\end{center}
\end{table}

The table shows that the use of the $2$-tree-lines do not find age-dependence that could not be found by the $1$-tree-lines. However, the ability of $2$-tree-lines to capture the two-split nature in the data sets is a clear advantage over $1$-tree-lines, and the computational ease of solving this problem presents this option as a valuable tool in searching for structure in tree data sets.

\section{Tree-Curves}\label{Treecurves}

A tree-curve is a sequence of trees, such that, given a tree in the tree-curve, the next tree in the sequence is obtained by adding one node. This node has to be a child of existing nodes in the previous tree to satisfy the connectivity requirement. The tree-curve idea is a generalization of the tree-line concept: the constraint on the location of the next added node is removed from the tree-line definition to obtain the definition of the tree-curve.

In Euclidean space, all points on a line are required to lie on a single direction. The constraint on the location of the next added node is considered to emulate this property in tree-lines. By removing it, a structure considered to be the counter part of a curve in Euclidean space is obtained.

\begin{definition}
A\textbf{\ tree-curve}, $C=\left\{c_{0},\cdots,c_{m}\right\}$, is a sequence
of trees where $c_0$ is called the starting tree, and $c_{i}$ comes from $c_{i-1}$ by the addition of a single
node, labeled $v_{i}$.
\end{definition}

\begin{figure}
[ptb]
\begin{center}
\includegraphics[
natheight=163pt,
natwidth=350pt,
height=163pt,
width=350pt
]%
{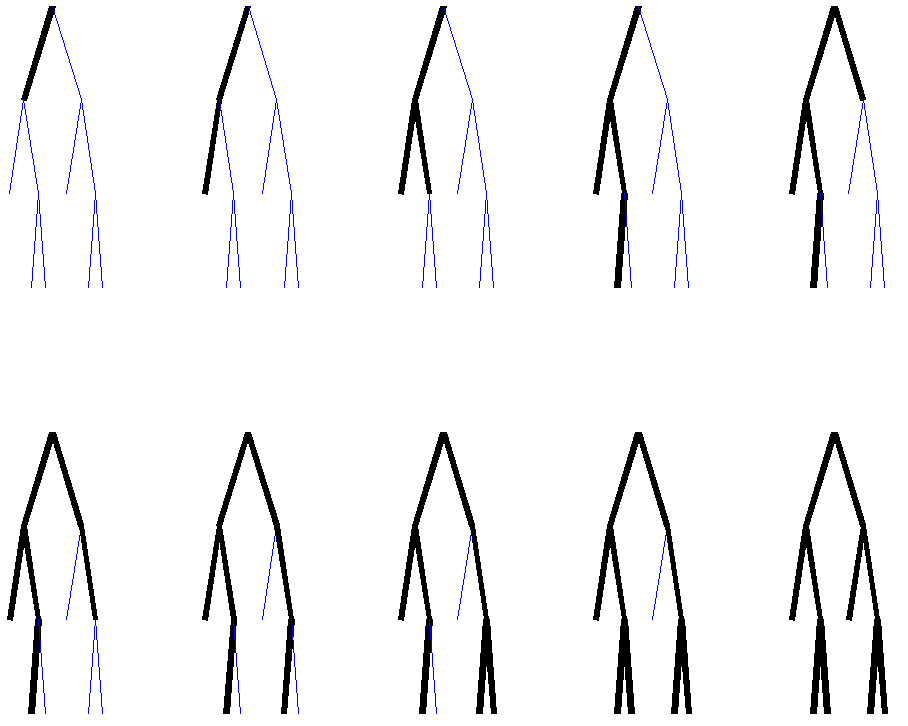}%
\caption{A toy example curve consisting of $10$ points. The initial tree is on the upper left. The curve finishes at the support tree on the lower right.}
\label{excurve}%
\end{center}
\end{figure}

An example tree-curve can be seen in Figure \ref{excurve}. Note that it starts from an initial tree of two nodes, and ends at the support tree.

The notions of projection and principal components for tree-curves follow what was introduced for $k$-tree-lines, with slight differences in notation.
 The projection of a data tree onto a tree-curve is the point on the tree-curve with smallest distance to the data tree:

\begin{definition}
Given a data tree $t$, its \textbf{projection} onto the tree-curve $C$ is%
\[
P_{C}\left(  t\right)  =\underset{c \in C}{\arg\min} \{ d\left(t,c \right)\}  .
\]
\end{definition}

The first principal component tree-curve is the curve that minimizes the sum of distances of each of the data points to their projections on the curve.

\begin{definition}
For a data set $T$, the \textbf{first principal component tree-curve} is
\[
C_1^* \, = \, \underset{C}{\arg\min} \sum_{t_i \in T} d(t_i,P_{C}(t_i))
\]
\end{definition}

The $j^{th}$ principal component tree-curve can be defined in a similar way. It will not be explicitly stated here since we do not provide methods to find them in this paper.


\subsection{Tree-Curve Solution Methods}\label{Cmethods}

Unlike the case with tree-lines, the sequence of nodes added to a starting point that define a tree-curve can be a member of a data tree in any order, as long as the connectivity requirement of the points on the tree-curve is satisfied. So far, this has prevented the development of an easy characterization of the projection of a data tree onto a tree-curve. Moreover, the set of all possible tree-curves on a given support tree has an order of $O(n!)$, where n is the number of nodes in the support tree.

We have not been able to solve the problem of finding the optimal first principal component to optimality. Given the very complex nature of this problem, it may be the case that the problem is NP-Hard. We developed some heuristic methods that give promising results. All heuristics mentioned below are known to give non-optimal results in some cases.

To test their effectiveness, a simulation with $30$ randomly generated data sets, each containing $4$ trees with $3$ levels, is run. This data set size is chosen so that the optimal best fitting tree-curve can be quickly found using an exhaustive search. The performance of each heuristic is measured by comparing their resulting tree-curve, $C$, with the optimal tree-curve $C^*$ that is found through exhaustive search. In particular, the performance of a tree-curve $C$ on a data set $T$ is measured using the objective function $F(C,T)$ value that needs to be minimized to reach the optimal tree-curve:

\[ F(C,T) =  \sum_{t_i \in T} d(t_i,P_{C}(t_i)) \]

And the performance percentage calculated is:

\[ \frac{F(C^*,T)}{F(C,T)}*100 \]

So far the following algorithms have been considered:

\subsubsection{Weight Order Algorithm (WO)}
This algorithm starts from a given starting tree, and adds the nodes from the support tree in the order of their weights (their number of occurrences in the data set). Ties are broken according to the parent-child relationship when possible: parents are added before their children. This algorithm achieved a performance measure of $98.82$. 

\subsubsection{Greedy Algorithm (G)}
Starting from an initial point, at each step the children of the existing nodes in the current step are considered. For each child, we calculate the improvement in the objective function if that node is added. The candidate with best contribution is appended to the current tree to obtain the next tree in the curve. This algorithm gave a performance of $89.76$.

\subsubsection{Switching Algorithm (S)}
This method starts from an arbitrary tree-curve, and considers pair of nodes that bring improvement in the objective function when their locations on the sequence defining the curve are switched. The method is terminated when no such pairs of nodes remain. When run using the original node order as a starting point, this algorithm performed at $94.02$.

\subsubsection{Weight Order + Switching Algorithm (WO+S)}
This method combines two of the heuristics mentioned above, by running the Weight Order algorithm first and feeding its result to the Switching algorithm, to see if any improvement can be achieved over the WO result by simple switching. This has proved to be the best performing method in the simulation with a measure of $99.91$, and is used to conduct the data analysis.

\subsection{Tree-Curve Data Analysis}\label{Cdata}
\begin{figure}
[ptb]
\begin{center}
\includegraphics[natheight=3in,natwidth=4.5in,height=3in,width=4.5in]
{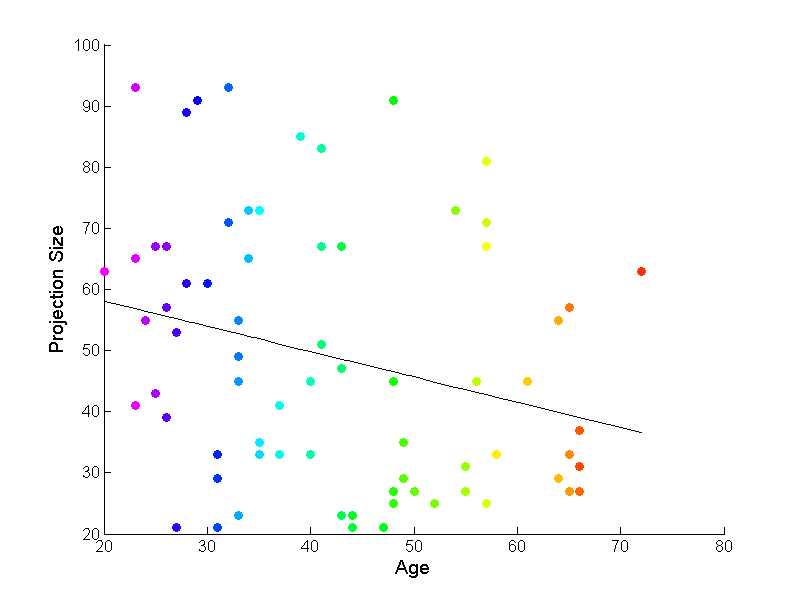}%
\caption{Size of projection onto tree-curve compared with age for left sub-population. The dots are colored according to age.}
\label{backcurve}%
\end{center}
\end{figure}

This data analysis has been conducted by running the WO+S method, since this one consistently gave the best results in our simulation. Each data point is projected onto the resulting best fitting tree-curve. Figure \ref{backcurve} shows an example of the relation between the size of this projection with the age of each subject. The black line was fitted to the data using linear regression. This plot was created for all of the sub-populations available, but this one is quite representative, so others are not shown to save space.

The tree-curve tool yields significant slope $p$-values for all of the sub-populations available in this data set. These are summarized in Table \ref{curvepvaltab}. The table contains the slope p-values obtained using the first principal $2$-tree-lines in Section \ref{sec:kanal}, and the first principal $1$-tree-lines in Ayd{\i}n et al. (2009). These results, together with further comparisons done using different versions of the brain artery data set and different correspondences can be found in Ayd{\i}n (2009) (see Tables $2.1$, $2.2$, $3.1$ and $4.2$). These strongly significant results obtained using tree-curves prove that this mode of analysis is a powerful tool to explain variation in binary trees.

\begin{table}[h]
\begin{center}
\begin{tabular}{| c | c | c | c | c |}
\hline
                 &Back   & Left   & Right  & Front \\ \hline
 $PC_{\infty}1$  &0.0285 & 0.0118 & 0.0246 & 0.0500  \\ 
 $PC_21$         &0.0159 & *      & *      & * \\
 $PC_11$         &0.0156 & *      & 0.0186 & *\\
\hline
\end{tabular}
\caption{The slope $p$-values obtained by the first principal tree-curve for all sub-populations (top row), in comparison with the results obtained by first principal $1$-tree-lines and $2$-tree-lines (next rows). The tree-curve p-values are all significant and and are overall better than what was found in previous work, showing the value of tree-curves. The $*$'s indicate p-values larger than $0.05$.}
\label{curvepvaltab}
\end{center}
\end{table}

In addition, as shown in Figure \ref{backcurve}, the very high projection sizes obtained renders this tool of analysis an attractive option. The first principal tree-curve captures $60\%$ of the nodes that exist in the data sets. This ratio again well exceeds what was obtained by the first principal $1$-tree-line ($12\%$) and the first principal $2$-tree-line ($16\%$). In fact, Ayd{\i}n et al. (2009) reports a $52\%$ coverage obtained by combining the first $10$ principal component $1$-tree-lines for descendant correspondence in their Figure $2.9$. The ability to summarize larger portions of data with the first principal component is a valuable contribution of tree-curves.

Note that for this tool, the length of the projection is not exactly equal to the number of nodes covered by the principal component, as was the case for tree-lines. Due to the structural nature of tree-curves, some nodes that do not exist in a data tree may appear in its projection.

A major drawback of tree-curves is the challenge of visually expressing the tree-curve resulting from an analysis run. Each tree-curve contains all the nodes that exist in the support tree, and what differentiates one curve from another is the sequence of nodes. Although it is possible to visually express a sequence in some ways (one can use changing rainbow colors, movies, etc.), visual inspection of those and trying to infer a structural trend from them is extremely hard. For example, structural properties observed using tree-lines (such as symmetry) are very challenging to infer from such visualizations.


\section{Discussion}

The statistical analysis of nontraditional data objects, such as shapes, images and graphs is a newly emerging and exciting area. This paper focuses on the analysis of populations of binary trees as data. The effort to develop principal component analysis tools using a combinatorial approach spans various papers in the literature: Wang and Marron (2007), Ayd{\i}n et al. (2009) and Alfaro et al. (2011). These studies developed various aspects of tree-line PCA, and reported promising numerical analysis results. Our paper generalizes the tree-line of the previous papers to $k$-tree-lines, providing a theoretical basis for a richer set of PCA tools capable of explaining various structures.

There are two special cases that we provide explicit tools for: $2$-tree-lines and tree-curves. Finding optimal $2$-tree-lines is shown to be a polynomial time problem. A new algorithm to improve the run times and memory requirements is given. There is an important property of a $2$-tree-line that enabled the $B\&B$ algorithm. When the projection of a data tree onto a $2$-tree-line is sought, there is a closed form expression that can be used to identify the projection. Such an expression also exists for $1$-tree-lines, enabling a linear-time algorithm. $B\&B$ leverages this expression to calculate the bounds on the candidate $2$-tree-lines without going over all the options. When $k>2$, there does not exist such an expression. This makes finding the $k$-tree-lines with $k>2$ difficult.

The numerical analysis results show that $2$-tree-lines are able to capture the double-branching nature of our data set, which tree-lines were not able to do due to shape limitations. The tree-curves prove to be a very powerful tool due to their flexibility to represent a variety of branching structures. This flexibility also brings computational difficulties. In this study we were able to find useful heuristics, but not a method guaranteed to find the actual optimum. Nevertheless, the application of the heuristics to our brain artery data set proves the representative power of tree-curves. The task of either finding a polynomial-time algorithm to find the optimal tree-curves, or proving that the problem is NP-hard is a future task.

All of the PCA tools for binary trees proposed in the literature so far assume that the analyst provides a suitable starting point for the $k$-tree-lines to grow from. This restriction can be lifted in future work, allowing a formulation where finding an optimal starting point becomes part of the problem.

In this framework, it is assumed that all nodes are identical: The only property distinguishing a node from another is its location. It is possible to construct a system where nodes carry other information as well. Wang and Marron (2007) formalize this idea where nodes have \emph{attributes}, and they develop a theoretical basis for these. However they do not provide a practical method to apply these in large scale data sets. The question of how to handle data sets with attributes is future work.

All of the above mentioned studies focus on PCA tools for trees. The area of developing methods to do classification is yet untouched. Such methods can have wide uses in actual data sets.

\section{Appendix}\label{Sec:2line}

\subsection{Proof of Lemma \ref{cl:2lineorder}}\label{lemmaproof}

The approach taken here is to count all possible $2$-tree-lines on a given data set. A polynomial bound on this number will suffice to conclude that we have a problem that can be solved in polynomial time, as the process of calculating the total distance of a given $2$-tree-line to the points in the data set is a linear-time process.

For a given data set, the number of possible $k$-tree-lines depends on the size of its support tree only, and not on the number of data trees in it. In this section, it will be assumed that the support tree is a full tree, i.e. all levels of the support tree include all the nodes on those levels. Another simplification is that we will assume the starting tree for the $2$-tree-lines considered is the root node. This approach will give an upper bound on the $2$-tree-line count, since arranging the same number of nodes in a full tree and starting from the root node would give the highest number of possible $2$-tree-lines. These two assumptions will enable us to disregard the structure of an arbitrary starting tree and the support tree in finding an upper bound that depends on the node count only.

Let: \\
$f(n)$ $=$ Number of $2$-tree-lines of which last added node is on $n^{th}$ level on a full support tree. \\
$f_1(n)$ $=$ Number of $2$-tree-lines in $f(n)$ with only one node on $n^{th}$ level\\
$f_2(n)$ $=$ Number of $2$-tree-lines in $f(n)$ with two nodes on $n^{th}$ level

We know that:
\[f(n) = f_1(n)+f_2(n) \ \ \ \forall n \geq 0\]

We will write a recursive formula for $f(n)$. If we consider the most trivial case where our tree is only the root node, we obtain the initial condition for the recursion:
\[f_1(1)=1\]
\[f_2(1)=0\]

To get the recursive formula, assume that we know the values of $f_1(n)$ and $f_2(n)$, and we are looking for $f_1(n+1)$ and $f_2(n+1)$. First let us count the $2$-tree-lines that end at $(n+1)^{st}$ level with a single node. This single node can be either one of the two children of a $2$-tree-line ending at level $n$ with a single node, or it can be one of the four children of a $2$-tree-line ending at level $n$ with two nodes. Therefore:
\[f_1(n+1)=2f_1(n)+4f_2(n)\]

For $f_2(n+1)$, first consider $f_1(n)$. These lines end with a single node at $n^{th}$ level, and have two children, where both of of them need to be added. Since the order of the addition matters, each such line gives us two options for extension. For $f_2(n)$, we need to choose two nodes out of the four children of $n^{th}$ level nodes. However, not all of the $2$-combinations of these are available. Now let us name the nodes on $n^{th}$ level as $a$ and $b$, $b$ being the last added node. Let us name their children as $a_1,a_2$ and $b_1,b_2$ respectively. Now the possible choices for addition are $(a_1,a_2)$, $(a_2,a_1)$, $(b_1,a_1)$, $(b_1,a_2)$, $(b_2,a_1)$, $(b_2,a_2)$. Summing all the choices up, we get:
\[f_2(n+1)=2f_1(n)+6f_2(n)\]

These two formulas are valid for all n greater than 1. Now let us write these two in matrix form:

\[  \left[ \begin{array}{ccc} f_1(n+1) \\ f_2(n+1) \end{array} \right] = \left[ \begin{array}{ccc}
                                                                   2 & 4 \\
                                                                    2 & 6 \end{array} \right] \left[ \begin{array}{ccc} f_1(n) \\ f_2(n) \end{array} \right] \]

Using this formula, we can write:
\[  \left[ \begin{array}{ccc} f_1(n+1) \\ f_2(n+1) \end{array} \right] = {\left[ \begin{array}{ccc}
                                                                   2 & 4 \\
                                                                    2 & 6 \end{array} \right]}^{n} \left[ \begin{array}{ccc} f_1(1) \\ f_2(1) \end{array} \right] \]

To further simplify this, we can re-write the coefficient matrix using spectral decomposition:
\[  \left[ \begin{array}{ccc} 2 & 4 \\ 2 & 6 \end{array} \right] = {\left[ \begin{array}{ccc}
                                                                   \sqrt{3}-1 & 1 \\
                                                                     1& \frac{1-\sqrt{3}}{2} \end{array} \right]}
                                                                   {\left[ \begin{array}{ccc} \lambda_1 & 0 \\ 0 & \lambda_2 \end{array} \right] }
                                                                   {\left[ \begin{array}{ccc}
                                                                   \sqrt{3}-1 & 1 \\
                                                                     1& \frac{1-\sqrt{3}}{2} \end{array} \right]}^{-1}
                                                                   \]

Where $\lambda_1 = 4+2\sqrt{3}$ and $\lambda_2 = 4-2\sqrt{3}$, the eigenvalues of coefficient matrix. Now we can get the $n^{th}$ multiple of this easily:
\[  {\left[ \begin{array}{ccc} 2 & 4 \\ 2 & 6 \end{array} \right]}^{n} = {\left[ \begin{array}{ccc}
                                                                   \sqrt{3}-1 & 1 \\
                                                                     1& \frac{1-\sqrt{3}}{2} \end{array} \right]}
                                                                   {\left[ \begin{array}{ccc} \lambda_1 & 0 \\ 0 & \lambda_2 \end{array} \right] }^{n}
                                                                   {\left[ \begin{array}{ccc}
                                                                   \sqrt{3}-1 & 1 \\
                                                                     1& \frac{1-\sqrt{3}}{2} \end{array} \right]}^{-1}
                                                                   \]

Insert this into $f(n+1)$ formula along with the initial conditions, and do the necessary simplifications:

\[ f_1(n+1) = \frac{\sqrt{3}-1}{2\sqrt{3}}(\lambda_1)^n + \frac{\sqrt{3}+1}{2\sqrt{3}}(\lambda_2)^n  \]

\[ f_2(n+1) = \frac{1}{2\sqrt{3}}(\lambda_1)^n - \frac{1}{2\sqrt{3}}(\lambda_2)^n  \]

Summing these up, we get the desired quantity:

\[f(n+1) = f_1(n+1)+f_2(n+1) = \frac{(\lambda_1)^n + (\lambda_2)^n}{2} \]

We know that a full support tree with $n$ levels has $\sim2^n$ nodes. If we call the total number of nodes in the support tree $m$, we have $n=\log_2(m)$. So for a problem with full support tree size $m$, the total number of $2$-tree-lines is:

\[ \frac{\lambda_1^{(\log_2m) - 1} + \lambda_2^{(\log_2m) - 1}}{2} \]

So the order of the problem of finding all $2$-tree-lines is:

\[ O(\frac{1}{2\lambda_1}m^{\log_2\lambda_1}) = O(m^{2.9})\]  

\subsection{Proof of Theorem \ref{cl:2linethm}}\label{thmproof}

Lemma \ref{cl:2lineorder} already establishes the order for the total count of $2$-tree-lines. To prove Theorem \ref{cl:2linethm}, we also need the maximum length of these $2$-tree-lines.

The full support tree with $m$ nodes has a depth $\log_2(m+1)$. And it is easy to see that the $2$-tree-line with maximum number of nodes in it that can be defined on this support tree has $1+2*(\log_2(m+1))$ nodes. We obtain this number by using the observation that a $2$-tree-line starting from the root can contain at most $2$ nodes from each level, except the root level. Therefore, the maximum number of nodes contained in each $2$-tree-line has an order of $O(\log m)$. Combine this with Lemma \ref{cl:2lineorder}, and we see that the order of all nodes contained in the list of all $2$-tree-lines is $O(m^{2.9}\log m)$.  The final step is to show that the brute force method needs to account every node on the $2$-tree-line list only once to form the list. This step is rather trivial, so it will not be elaborated here.

\subsection{Proof of Proposition \ref{lowerbound}}

The projection of a data point onto an object is, naturally, a point on that object. In our case, this implies the fact that the projection of a data point $t_i$ onto a $2$-tree-line $K$, $P_K(t_i)$, is a tree that is contained in $Pa(K)$. Therefore we can write:
\begin{eqnarray}\label{bir}
Pa(K)&\supseteq& P_K(t_i) \notag \\
t_i\cap Pa(K) &\supseteq& t_i \cap P_K(t_i) \notag \\
|t_i\cap Pa(K)| &\geq & |t_i \cap P_K(t_i)|
\end{eqnarray}

Let $K^*$ be any maximal $2$-tree-line that can be extended from $K$. Naturally, $K^*\supseteq K$, and:

\begin{equation}\label{iki}
\sum_{v \in MP(K)}{w(v)} \geq \sum_{v \in Pa(K^*)}{w(v)}
\end{equation}

Now,  using (\ref{bir}) and (\ref{iki}), we can show:

\begin{eqnarray*}
\sum_{t_i \in T} d(t_i,P_{K^*}(t_i)) &=& \sum_{t_i \in T} (|t_i| - |t_i \cap P_{K^*}(t_i)| + |P_{K^*}(t_i) \backslash t_i|) \\
 &=& \sum_{t_i \in T}{|t_i|} - \sum_{t_i \in T}{|t_i \cap P_{K^*}(t_i)|} + \sum_{t_i \in T}{|P_{K^*}(t_i) \backslash t_i|}\\
 &\geq & \sum_{t_i \in T}{|t_i|} -\sum_{t_i \in T}{|t_i \cap P_{K^*}(t_i)|}\\
 &\geq& \sum_{t_i \in T}{|t_i|} - \sum_{t_i \in T}{|t_i \cap Pa(K^*)|}\\
 &=&\sum_{t_i \in T}{|t_i|} - \sum_{v \in Pa(K^*)}{w(v)}\\
 &\geq &\sum_{t_i \in T}{|t_i|} - \sum_{v \in MP(K)}{w(v)}\\
\end{eqnarray*}

Which proves that any maximal $2$-tree-line extending from $K$ will have a worse objective function value than $\sum_{t_i \in T}{|t_i|} - \sum_{v \in MP(K)}{w(v)}$, and therefore $\sum_{v \in MP(K)}{w(v)}$ provides a lower bound.


%
%

\section{Acknowledgements}
During this research,  Burcu Ayd{\i}n was partially supported by NSF grants DMS-0606577 and DMS-0854908, and NIH Grant
RFA-ES-04-008. Haonan Wang was partially supported by NSF grants DMS-0706761 and DMS-0854903.
Alim Ladha and Elizabeth Bullitt were partially supported by NIH grants R01EB000219-NIH-NIBIB and R01 CA124608-NIH-
NCI.
J.S. Marron was partially supported by NSF grants DMS-0606577 and DMS-0854908, and NIH Grant
RFA-ES-04-008.

The final publication of this paper will be available at springerlink.com, in Statistics and Biosciences journal.


\begin{thebibliography}{}
%
%
\bibitem{Alfaro2011} Alfaro, C.A., Ayd{\i}n, B., Bullitt, E., Ladha, A., Valencia, C.E., Dimension Reduction in Principal Component Analysis for Trees, \textit{Submitted to Statistics and Computing.} (2011)

\bibitem{Aydin2009} Ayd{\i}n, B., Pataki, G., Wang, H., Bullitt, E., Marron, J.S., A Principal Component Analysis For Trees,\textit{Annals of Applied Statistics},  3:1597–1615 (2009)

\bibitem{Aydin2011} Ayd{\i}n, B., Pataki, G., Wang, H., Ladha, A., Bullitt, E., and Marron, J.S.,  Visualizing the
Structure of Large Trees, \textit{Electronic Journal of Statistics}, Volume 5, 405-420 (2011)

\bibitem{AydinPHD2009} Ayd{\i}n, B., Principal Component Analysis of Tree Structured Objects, \textit{Ph.D. Thesis}, University of North Carolina at Chapel Hill. (2009)

\bibitem{banks1998} Banks, D. and Constantine, G. M., Metric Models for Random Graphs, \textit{J.
Classification} 15 199-223 (1998)

\bibitem{bazaraa1979} Bazaraa, M. S. and Shetty, C. M., Nonlinear programming: Theory and algorithms, \textit{John Wiley and Sons} (1979)

\bibitem{bullitt2002}Aylward, S. and Bullitt, E., Initialization, Noise, Singularities and Scale in Height Ridge Traversal for Tubular Object Centerline Extraction, \textit{IEEE Transactions on Medical Imaging}, 21, 61-75 (2002)

\bibitem{bullitt2003} Bullitt, E., Gerig, G., Pizer, S.M., Aylward, S.R., Measuring tortuosity of the intracerebral vasculature from
MRA images, \textit{IEEE Transactions on Medical Imaging}, 22, 1163-1171  (2003)

\bibitem {bullitt2010}Bullitt, E., Zeng, D., Ghosh, A., Aylward, S. R., Lin, W., Marks, B. L., Smith, K., The Effects of Healthy Aging on Intracerebral Blood Vessels Visualized by Magnetic Resonance Angiography, \textit{Neurobiology of Aging}, 31(2), 290–300 (2010)

\bibitem{breiman1984} Breiman, L., Friedman, J. H., Olshen, J. A., Stone, C. J., \textit{Classification
and Regression Trees} Belmont, CA: Wadsworth (1984)

\bibitem{breiman1996}  Breiman, L., Bagging Predictors, \textit{Machine Learning}, vol 24, Number 2,
123-140 (1996)

\bibitem{cook1997} Cook, W. J., Cunningham, W. H., Pulleyblank, W. R., Schrijver, A., Combinatorial Optimization, \textit{John Wiley and Sons} (1997)

\bibitem{everitt2001} Everitt, B. S., Landau, S., Leese, M., \textit{Cluster Analysis (4th edition)},
Oxford University Press, New York (2001)

\bibitem {handle} Handle, http://hdl.handle.net/1926/594 (2008)

\bibitem{land1960}Land, A. H. and Doig, A. G., An Automatic Method of Solving Discrete Programming Problems, \textit{Econometrica} 28 (3), pp. 497-520 (1960)

\bibitem{lawlerwood1966} Lawler, E. L. and Wood, D. E., Branch-and-bound methods: A survey, \textit{Operations Research}, 14, 699–719 (1966)

\bibitem{lawlerbell1966} Lawler, E. L. and Bell, M. D., A Method for Solving Discrete Optimization Problems, \textit{Operations Research}, Vol. 14, No. 6, pp. 1098-1112 (1966)

\bibitem{nye2011}  Nye,T., Principal Component Analysis in the Space of Phylogenetic Trees, Unpublished Manuscript, http://www.mas.ncl.ac.uk/~ntmwn/pca/preprint.pdf (2011)

\bibitem{Schrijver1998} Schrijver,A., Theory of linear and integer programming, \textit{John Wiley and Sons} (1998)

\bibitem{shen2011}	Shen, D., Shen, H.,  Bhamidi, S.,  Munoz-Maldonado,Y., Kim, Y., Marron, J.S. Functional Data Analysis for Trees. Manuscript in progress. (2011)

\bibitem{wang2007}Wang, H. and Marron, J.S., Object Oriented Data Analysis: Sets of Trees, \textit{Annals of Statistics}, 35, 1849-1873 (2007)

\bibitem{wang2011} Wang,Y.,  Marron, J.S., Ayd{\i}n, B.,  Ladha, A., Bullitt, E. and Wang,H., Nonparametric Regression Model with Tree-structured Response, submitted to \textit{JASA Case Study}. (2011)



\end{thebibliography}


\end{document}